\documentclass[cpp,a4paper,fleqn%
]{w-art}
\usepackage{times,cite,w-thm,amsmath,amssymb,url}

\theoremstyle{plain}

\theoremstyle{definition}

\usepackage[]{graphicx}
\begin{document}
\DOIsuffix{theDOIsuffix}
\Volume{46}
\Month{01}
\Year{2007}
\pagespan{1}{}
\Receiveddate{XXXX}
\Reviseddate{XXXX}
\Accepteddate{XXXX}
\Dateposted{XXXX}



\title[Ab initio simulation of complex dielectric function]{Ab initio simulation of complex dielectric function for dense aluminum plasma}


\author[M.E. Povarnitsyn]{M. E. Povarnitsyn\inst{1}}
\author[D.V. Knyazev]{D. V. Knyazev\inst{1,2}}
\author[P.R. Levashov]{P. R. Levashov\inst{1,2}
  \footnote{Corresponding author\quad E-mail:~\textsf{pasha@ihed.ras.ru},
            Phone: +7\,495\,484\,24\,56,
            Fax: +7\,495\,485\,79\,90}}
\address[\inst{1}]{Joint Institute for High Temperatures, Izhorskaya 13 bldg 2, Moscow 125412, Russia}
\address[\inst{2}]{Moscow Intitute of Physics and Technology (State University), Institutsky lane 9, Dolgoprudny, Moscow region 141700, Russia}
\begin{abstract}
We present calculations of frequency-dependent complex dielectric function of dense aluminum plasma by quantum molecular dynamics method for temperatures up to
20~kK. Analysis shows that the dependencies for real and imaginary parts can be interpolated by the Drude formula with two effective
parameters: the mean charge of ions and the effective frequency of collisions. The rise of these parameters with temperature deviates from simple theoretical predictions.

\end{abstract}
\maketitle                   





\section{Introduction}

The frequency-dependent complex dielectric function of metals is a macroscopic parameter which characterises their optical properties
under different conditions. Interaction of electromagnetic radiation with matter can be computed from the solution of the Maxwell equations with the closure given by the dielectric function \cite{Taflove:2000}. For this value there are two well-known cases for theoretical approaches: the metallic plasma with
nearly-free electrons below the Fermi-temperature \cite{Ashkroft:1976, Born:1980} (the Drude theory), and the weakly-coupled non-degenerate plasma \cite{tvt03, PoP06}. The first case is valid at near-normal densities and relatively low temperatures, the second one---at low densities and high temperatures.
The intermediate region of parameters is hard-to-reach both for theory and computations, so the usual approach is to
interpolate between these two cases \cite{Basko:PRE:1997, Eidmann:PRE:2000, Veysman:JEPT:2007}.
%
%

During the last decade the computations of the dielectric function of metals by the quantum molecular dynamics (QMD) and Kubo-Greenwood formalism became available at temperatures below the Fermi-temperature \cite{Desjarlais:PRE:2002, Mazevet:PRE:2005}.
These calculations give information both for the static electrical conductivity and optical properties of dense metallic plasma. In this work we analyse the QMD-dependencies of complex dielectric function for dense aluminum plasma at temperatures up to 20~kK. This is important for a
calibration of semiempirical models of complex dielectric function which are necessary for the simulation of laser-matter interaction.

\section{Ab initio calculations}
Ab initio calculations of the complex dielectric function consist of three main stages: QMD simulation, precise zone structure
resolution and further calculation of the dynamical electrical conductivity and complex dielectric function.

QMD simulation of an atomic supercell is performed in the framework of the density functional theory (DFT). At each ionic step the electronic ground state is calculated within the Born--Oppenheimer approximation: due to the great difference in masses between electrons and ions, electrons immediately adjust to the current ionic configuration. Core electrons are treated within a pseudopotential approach, that significantly reduces the computational time. In our calculations we use ultrasoft pseudopotentials (US-PP) of Vanderbilt \cite{Vanderbilt:PRB:1990}. This speeds up the QMD simulation in comparison with projector augmented wave (PAW) pseudopotentials with insignificant influence on the results. Ions are treated classically. The well-known Verlet algorithm is applied to integrate the Newton equations of ionic motion with the Hellmann-Feynman forces acting on ions. QMD method is implemented in Vienna ab initio simulation package (VASP) \cite{Kresse:PRB:1993, Kresse:PRB:1994, Kresse:PRB:1996}. Simulations were performed with 108 atoms in the supercell during 1500 steps (each time step was equal to 2~fs). At the beginning of QMD simulation the ions are located in the fcc lattice. The simulation in the liquid phase consists of two sections: at the first section the initial ionic configuration is destroyed and the energy tends to some equilibrium value (this requires about 100 steps); at the second, equilibrium section, energy fluctuates around its equilibrium value. The number of steps must be large enough to get several independent ionic configurations from the equilibrium section of MD run. Following \cite{Desjarlais:PRE:2002}, $\Gamma$-point only was used during the electronic ground-state calculation. Number of bands varied from 300 at 3~kK up to 1000 at 20~kK at normal density. Energy cut-off was taken as 100~eV.

Independent ionic configurations, chosen from the equilibrium section of MD run, are used for better resolution of a zone structure. In comparison with QMD simulation bigger numbers of bands, $\mathbf k$-points in the Brilloiun zone (BZ) and energy cut-off can be taken. Electronic wave functions obtained by solution of the Kohn-Sham (KS) equations are further used for the calculation of the electrical conductivity. During our calculation 15 ionic configurations were chosen, the statistical error in this case was less than 2\%. Number of bands varied from 600 at 3~kK up to 1300 at 20~kK at normal density. The calculations were performed both for $\Gamma$ point only and denser meshes in the BZ according to the Monkhorst-Pack scheme \cite{Monkhorst:PRB:1976}. In the liquid phase the results  turned out to be practically insensitive to the number of $\mathbf k$-points in the BZ. Energy cut-off was the same as during the QMD simulation, $E_{\text{cut}}= 100$~eV.

For each ionic configuration the real part of dynamical electrical conductivity for the frequency $\omega$ is calculated using the Kubo-Greenwood formula \cite{Moseley:AJP:1978}:
\begin{equation}
\sigma_1\left(\omega\right) =
\frac{2\pi e^2 \hbar^2}{3m^2 \omega \Omega}\sum_{\mathbf k, i, j, \alpha} W(\mathbf k)\left[F\left(\epsilon_{i,\mathbf k}\right)-F\left(\epsilon_{j,\mathbf k}\right)\right]\left|\left<\Psi_{j,\mathbf k}\left|\nabla_\alpha\right|\Psi_{i,\mathbf k}\right>\right|^2 \delta\left(\epsilon_{j,\mathbf k} - \epsilon_{i,\mathbf k} - \hbar\omega\right),
\end{equation}
where $e$, $m$, $\hbar$ are the electron charge, the electron mass and the reduced Planck constant, respectively; $\Omega$ is the volume of the supercell under consideration. The sum is over all $\mathbf k$-points within the BZ (by $\mathbf k$), all bands participating in the calculation (by $i$ and $j$), and three spatial dimensions (by $\alpha$). $W\left(\mathbf k\right)$ is the weight of the particular $\mathbf k$-point in the BZ, $\epsilon_{i,\mathbf k}$ is the energy of the band number $i$ for the given $\mathbf k$-point, $F\left(\epsilon_{i,\mathbf k}\right)$ is the Fermi-weight of this band, $\Psi_{i,\mathbf k}$ is the corresponding wave function, received as the solution of the KS equations. Unfortunately, at this stage only pseudo wave functions may be used in the conductivity calculation and the correction term for US-PP \cite{Desjarlais:PRE:2002} is not taken into account.
Nevertheless, our simulation of optical conductivity of aluminum at $\rho = 2$~g/cm$^3$ and $T = 10$, 20~kK is in very good agreement with the results obtained with the PAW pseudopotential \cite{Desjarlais:PRE:2002}.
Dynamical electrical conductivity is calculated in the range of frequencies from 0.02 up to 10~eV. For high-frequency values of conductivity to be correct, large enough number of bands should be taken during the precise resolution of the electronic structure. The $\delta$-function is broadened during the calculation with the Gaussian function. Following \cite{Desjarlais:PRE:2002} the broadening must be as small as possible, but not too small to prevent unphysical oscillations due to the discreteness of the electronic states. The value of the broadening of 0.1~eV (the standard deviation of the Gaussian) was chosen for the liquid phase at normal density. The values of dynamical electrical conductivity for different ionic configurations are averaged.

With real part of conductivity known, imaginary part is derived using the Kramers-Kronig relation:
\begin{equation}
\sigma_2\left(\omega\right)=-\frac{2}{\pi}P\int_0^\infty\frac{\sigma_1\left(\nu\right)\omega}{\nu^2-\omega^2}d\nu,
\end{equation}
where $P$ designates the principal value of the integral \cite{Collins:PRB:2001}. The real and imaginary parts of the dielectric function $\varepsilon=\varepsilon_1+i\varepsilon_2$ are calculated via the formulas (SI)
$\varepsilon_1\left(\omega\right)=1-\frac{\sigma_2\left(\omega\right)}{\omega\varepsilon_0}$,
$\varepsilon_2\left(\omega\right)=\frac{\sigma_1\left(\omega\right)}{\omega\varepsilon_0}$,
where $\varepsilon_0$ is the vacuum permittivity.

\begin{figure}[htb]
\centering\includegraphics[width=0.45\columnwidth]{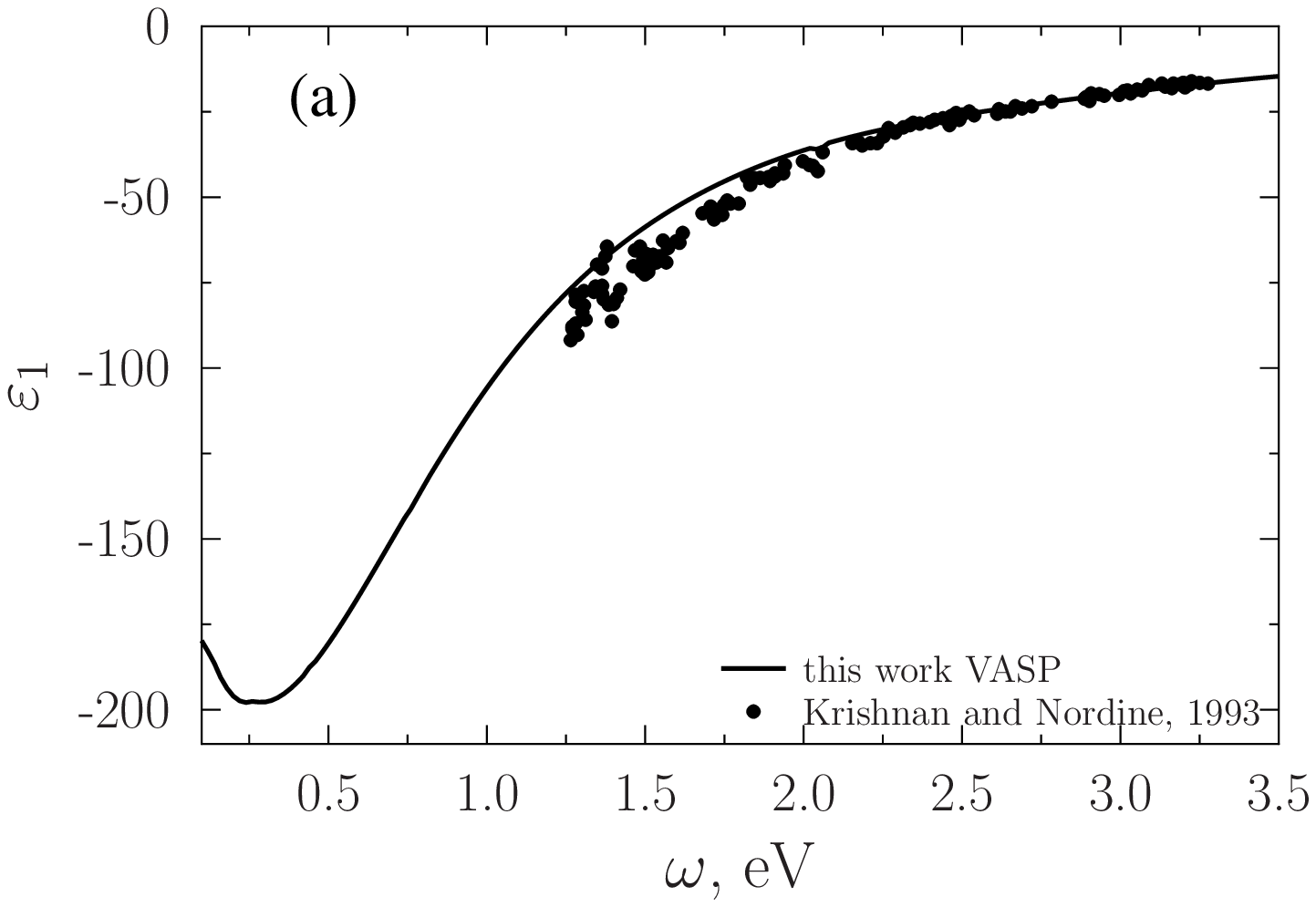}
\includegraphics[width=0.45\columnwidth]{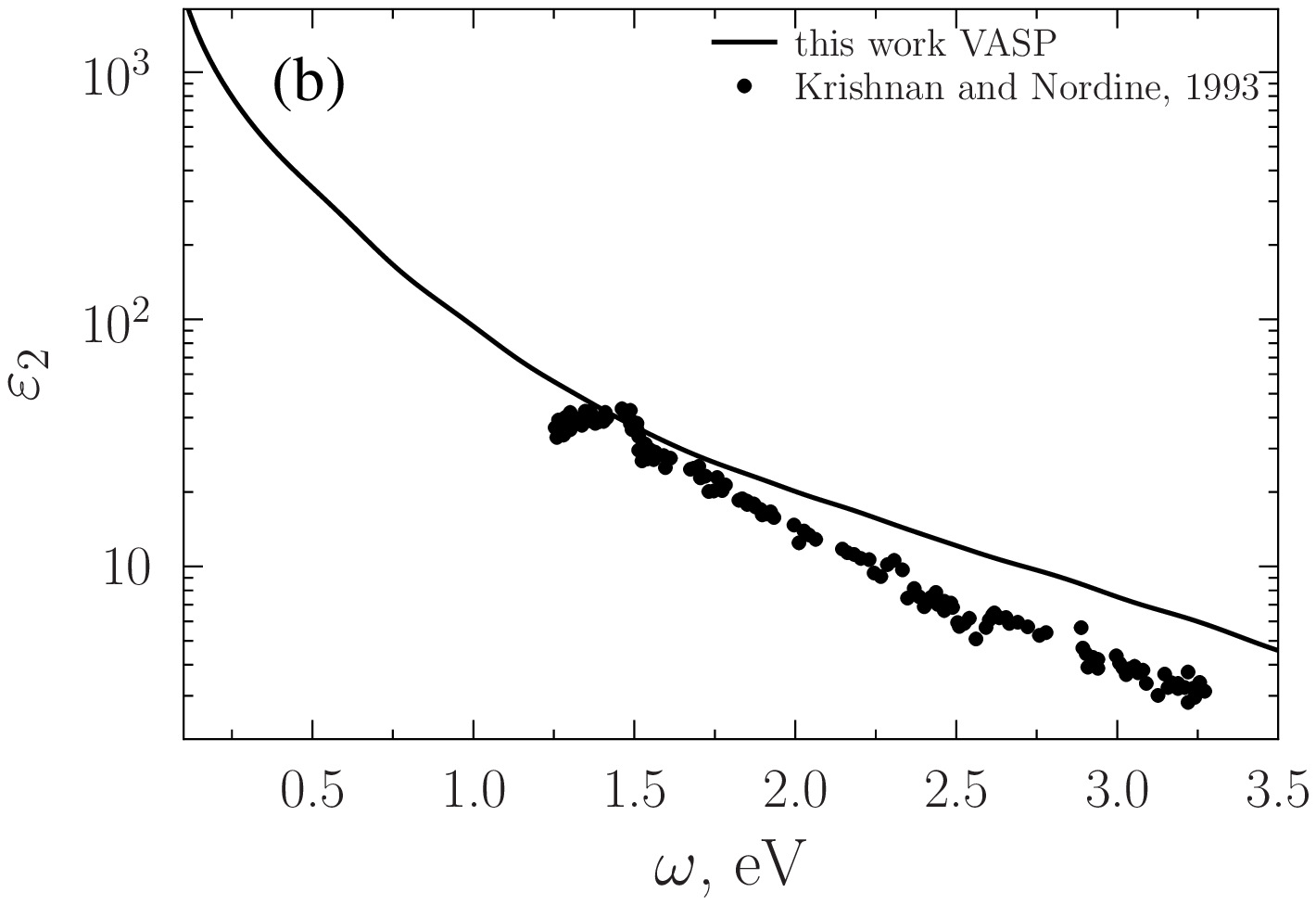}
\caption{The real (\textbf{a}) and imaginary (\textbf{b}) parts of dielectric function vs. frequency for aluminum at $T=1550$~K and $\rho=2.23$~g/cm$^3$. Solid line---this work (VASP calculations), signs---experimental data \cite{Krishnan:PRB:1993}.}
\label{Al_1550K_223gcm3}
\end{figure}

The results of the calculations for aluminum at $T=1550$~K, $\rho=2.23$~g/cm$^3$ and available experimental data \cite{Krishnan:PRB:1993} are shown in Fig.~\ref{Al_1550K_223gcm3}.  The curves of dynamical electrical conductivity and dielectric function have the Drude-like shapes. The results of the calculations are in the satisfactory agreement with experimental data, this fact confirms that our computational scheme is reliable.

\begin{figure}[htb]
\centering\includegraphics[width=0.45\columnwidth]{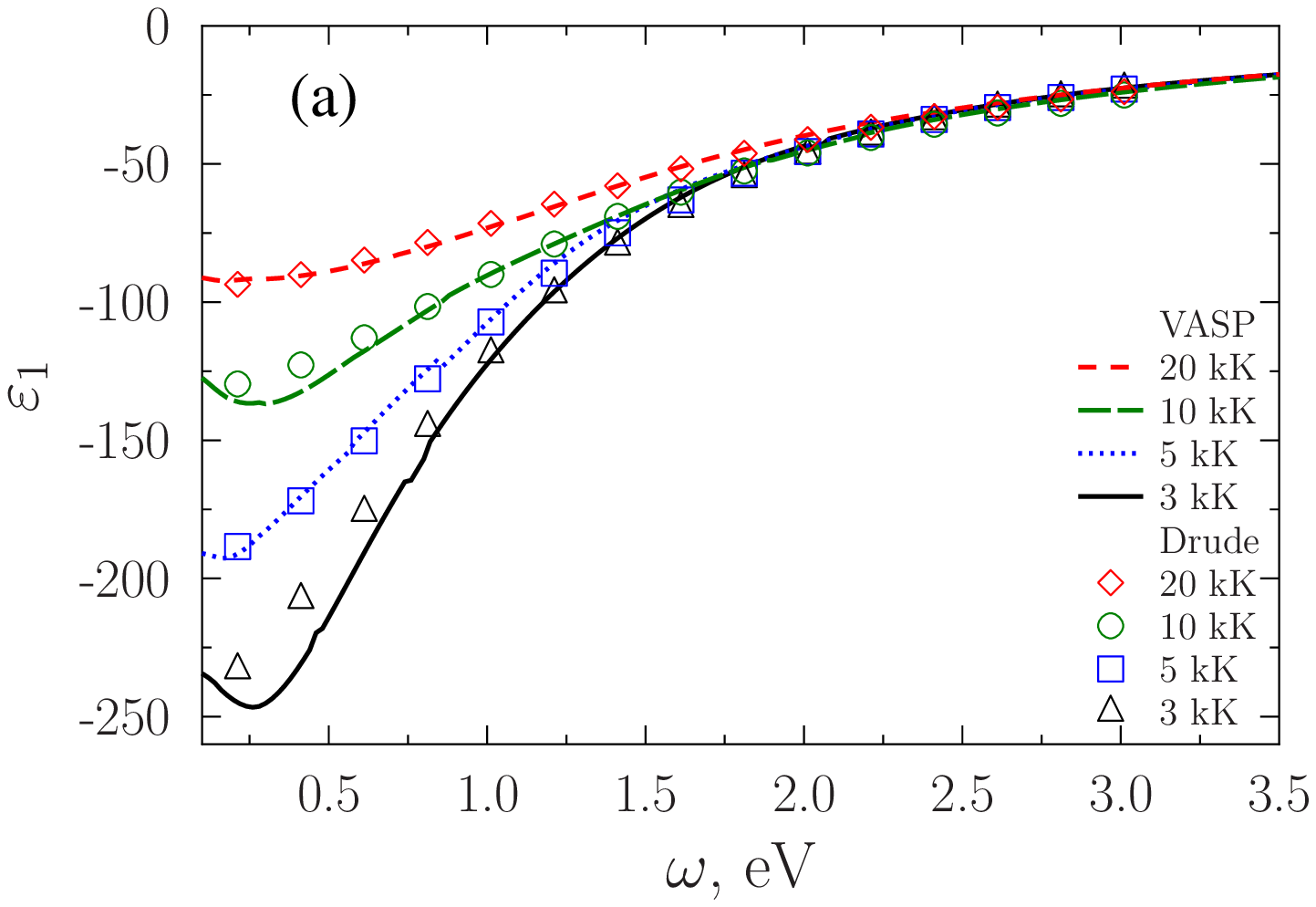}
\includegraphics[width=0.45\columnwidth]{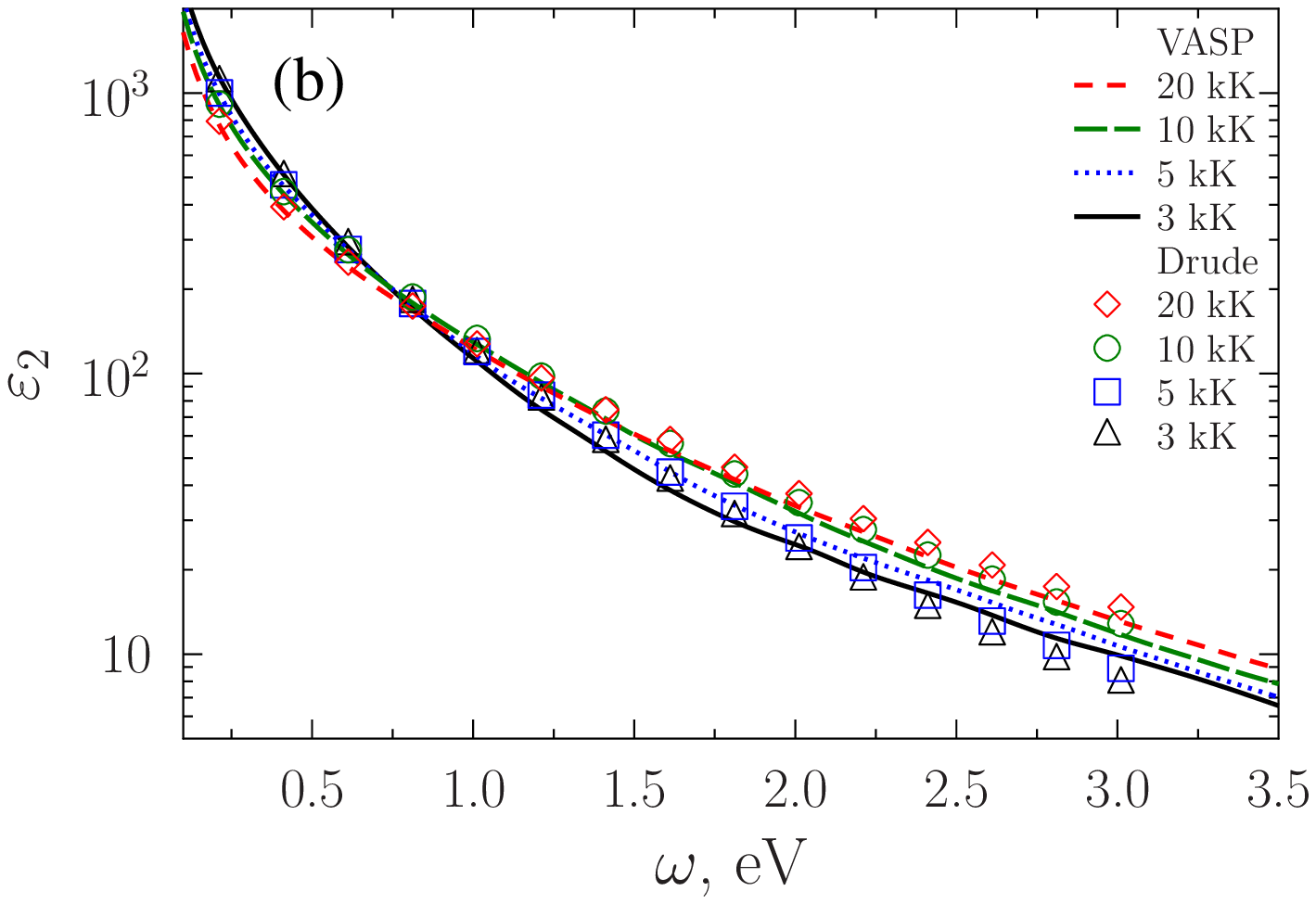}
\caption{The real (\textbf{a}) and imaginary (\textbf{b}) parts of dielectric function for aluminum at normal isochor ($\rho=$2.70~g/cm$^3$) for 3, 5, 10 and 20~kK. Curves---VASP calculations, markers---Drude model.}
\label{Al_normal_isochor}
\end{figure}

The real and imaginary parts of the complex dielectric function along the normal isochor  at temperatures from 3 to 20~kK are shown in Fig.~\ref{Al_normal_isochor}.


\section{Discussion}

It can be clearly seen from Fig.~\ref{Al_normal_isochor} that at all temperatures the obtained dependencies of the complex dielectric function are Drude-like. The Drude model \cite{Ashkroft:1976,Born:1980} is determined by the formula $\varepsilon(\omega, n_e, T)=1 - \frac{\omega_\mathrm{pl}^2} {\omega(\omega + i\nu_{\mathrm{eff}})}$,
%
%
where $\omega_\mathrm{pl} = \sqrt{4\pi n_e e^2/m}$ is the plasma frequency with the electron concentration $n_e$, $\nu_{\text{eff}}$ is the effective collision frequency. The model contains two
parameters: $n_e = \left< Z\right> n_a$, where $n_a$---the concentration of ions, $\left< Z\right>$---the mean ion
charge, and the effective collision frequency $\nu_{\text{eff}}$. The result of the adjustment of these two
parameters to the QMD-dependencies of the complex dielectric function is also shown in Fig.~\ref{Al_normal_isochor}.



\begin{figure}[t]
\centering\includegraphics[width=0.48\columnwidth]{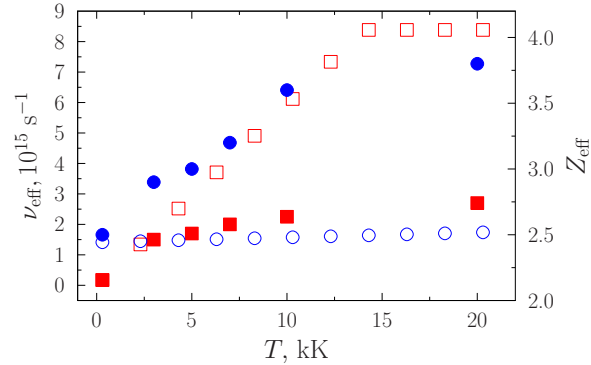}
\caption{Left axis: effective collision frequency vs. temperature at normal isochor of aluminum, empty squares---\cite{Povarnitsyn:ASS:2011}, filled squares---this work. Right axis:  mean ion charge vs. temperature at normal isochor for aluminum, empty circles---\cite{Povarnitsyn:ASS:2011}, filled circles---this work.}
\label{nu_and_z}
\end{figure}

The mean ion charge and the effective collision frequency obtained by the interpolation of QMD results to the
Drude formula and by other models are shown in Fig.~\ref{nu_and_z}.
It is clearly seen that the mean ion charge $\left< Z\right>$ rises significantly with temperature (filled circles), while the
estimation from the finite-temperature Thomas-Fermi model (empty circles) shows almost constant value in the range
3--20~kK.


The frequency of collisions in a metal state is a sum of electron-phonon and electron-electron collisions $\nu_\mathrm{eff}=\nu_\mathrm{e-ph}+\nu_\mathrm{ee}$ \cite{Eidmann:PRE:2000, Veysman:JEPT:2007, Povarnitsyn:ASS:2011}.
%
%
The first one is proportional to the ion subsystem temperature $\nu_\mathrm{e-ph}\sim T_i$; the second one---to the square of electron temperature $\nu_\mathrm{ee}\sim T_e^2/T_F$. Also the effective frequency of collisions is limited by the electron free-path between ions; for the normal density this limitation is valid for the model \cite{Povarnitsyn:ASS:2011} at $T\gtrsim 15$~kK. One can see from Fig.~\ref{nu_and_z} that the QMD collision frequency is a slowly increasing function of temperature (filled squares) while the model \cite{Povarnitsyn:ASS:2011} shows significant
rise with temperature (empty squares). The possible explanation of distinctions for the mean ion charge and effective collision
frequency in comparison with \cite{Povarnitsyn:ASS:2011} is the neglect of band-to-band
transitions in the Drude formula. The contribution of this effect is significant in solid aluminum \cite{Huttner}, however, currently it is not clear at what temperature the band-to-band transitions disappear. Probably
this term should be present in the Drude formula even at temperatures higher than the melting temperature, and subsequent QMD calculations might clarify this question.

\section{Conclusion and acknowledgements}
We have calculated the complex dielectric function of dense aluminum plasma along the normal isochor up to 20 kK by QMD. The analysis of the obtained dependencies has shown that despite the Drude-like character
of $\varepsilon_1$ and $\varepsilon_2$ additional contribution to the dielectric function should be taken into account. For solid aluminum the band-to-band transitions are significant, so it is possible that even at high
temperatures this process can not be neglected.


This work is supported by the Council of the President
of the Russian Federation for Support of Young Russian Scientists and
Leading Scientific Schools (project No. NSh-65792.2010.2), the
Ministry of Education and Science of the Russian Federation (the Federal
targeted program ``Research and development in priority fields of
scientific and technological complex of Russia'' 2007--2013, contract No. 07.514.12.4002), and the FRRC grant for master students (2011-2012).

\hyphenation{Post-Script Sprin-ger}
\providecommand{\WileyBibTextsc}{}
\let\textsc\WileyBibTextsc
\providecommand{\othercit}{}
\providecommand{\jr}[1]{#1}
\providecommand{\etal}{~et~al.}


\end{document}